# A TEMPERATURE-MAPPING SYSTEM FOR MULTI-CELL SRF ACCELERATING CAVITIES*

M. Ge, G. Hoffstaetter, F. Furuta, E. Smith, M. Liepe, S. Posen, H. Padamsee, D. Hartill, X. Mi, Cornell University, Ithaca, NY 14850, USA.

*Abstract*

A Temperature mapping (T-map) system for Superconducting Radio Frequency (SRF) cavities consists of a thermometer array positioned precisely on an exterior cavity wall, capable of detecting small increases in temperature; therefore it is a powerful tool for research on the quality factor ($Q_0$) of SRF cavities. A new multi-cell T-mapping system is has been developed at Cornell University. The system has nearly two thousand thermometers to cover 7-cell SRF cavities for Cornell's ERL project. A new multiplexing scheme was adopted to reduce number of wires. A 1mK resolution of the temperature increase $\Delta T$ is achieved. A 9-cell cavity of TESLA geometry was tested with the T-map system. By converting $\Delta T$ to power loss and quality factor, it has been found that for this cavity, most surface losses were generated by the first cell when the accelerating gradient is increased above 15MV/m. The comparison of Q-value between with and without hotspots shows the heating on cavity wall degraded cavity $Q_0$ about 1.65 times. The power loss on the hotspots is about 40% of the total power. Effective and intuitive ways of displaying surface properties of the cavity interior, e.g. the residual resistivity, will be shown.

## INTRODUCTION

The quality factor ($Q_0$) of Superconducting Radio Frequency (SRF) cavities is a critical topic in the SRF field. High-Q cavities are able to save cryogenic capacities of accelerators, therefore reduce total costs of accelerators. The definition of $Q_0$ is given by equation (1), where $\omega$ is the angular frequency of the cavity, $U$ is its stored energy, $P_{loss}$ is the power loss on the cavity surface. The $Q_0$-value ranks the cavity's efficiency by comparison of the energy stored and lost. The high $Q_0$-value which people are seeking indicates less loss on the surface, hence less surface heating, at the same stored energy level. If we define a geometry factor ($G$) as function (2), the $Q_0$ may be written as the ratio of the geometry factor and the surface resistance ($R_s$).

$$Q_0 = \frac{\omega U}{P_{loss}} = \frac{G}{R_s} \qquad (1)$$

$$G = \frac{\omega \mu_0 \int_V |\vec{H}|^2 dV}{\int_A |\vec{H}|^2 dA} \qquad (2)$$

The geometry factor is a constant determined by cavity geometry shapes, thus the $Q_0$ is dictated by cavity surface-resistance. In other words, the way to obtain a high-Q cavity is to reduce its surface resistivity which is determined by surface preparation of the cavity such as chemical etching, high-pressure water rinsing, and cleaning assembly *etc*.

The T-map system, a thermometer array uniformly attached on a cavity exterior surface, is a powerful tool for surface resistance research. It is able to detect tiny heating on cavity wall during feeding RF power into cavity. Hence by the heating map of whole cavity, it is possible to calculate power loss and surface resistance, furthermore to calculate the $Q_0$-value of the cavity. The $Q_0$ of a multi-cell SRF cavity measured from RF is derived from the total value of all cells. The relationship between total $Q_0$ of a 9-cell cavity and $Q_0$-value of each cell at pi-mode is given by equation (3)

$$\frac{1}{Q_0^{tot}} = \frac{1}{n}\sum_{n=1}^{9}\frac{1}{Q_n^{cell}} \qquad (3)$$

Here $Q_0^{tot}$ is total $Q_0$ of a 9-cell cavity, and $Q_n^{tot}$ is $Q_0$ of cell $n$. The total $Q_0$ is unable to reflect which cell has low-Q and degrades the whole cavity. However, T-map can diagnose the $Q_0$-value of each cell by summing each cell's power loss. Hence T-map gives more information about cavity performance than RF measurement alone. The useful information gives effective feedback on cavity surface preparation and guides the next step of cavity treatment.

T-mapping history can be traced back to the 1980s. Cornell University was a pioneer in developing a 1-cell T-map system for 1.5GHz SRF cavity research [1, 2, and 3]. Now Jefferson Lab and Fermilab also have 1-cell or 2-cell T-map systems. These systems are used for fundamental SRF research via single-cell cavities. However, the recipe of cavity treatment developed for single-cell cavities is difficult to transfer to multi-cell cavities which are the real components for accelerators. The yield of performance of multi-cell cavity is about 30% [4]. Therefore the development of a multi-cell T-mapping system is necessary and ultra-important. DESY and Los Alamos have multi-cell T-mapping systems for ILC 9-cell cavities [5, 6]. Those systems are mainly used to detect quench location.

The Cornell multi-cell T-mapping system, by virtue of its high sensitivity, is able to detect heating levels much lower than those required to cause a quench. The hot-spots normally start in the medium accelerating gradient region and cause a Q-drop. Increasing the accelerating gradient of cavity, the heating at a hot-spot will grow as well, and eventually quench the cavity. Therefore it is important to discover not only the quench location, but also the original location of heating as well as the growth

___________

*This work has been supported by NSF award PHY-0969959 and DOE award DOE/SC00008431

of the heating rate versus accelerating gradient. This information would help to unveil the loss mechanism of superconductivity under medium RF field, and address fundamental physics.

## TEMPERATURE-MAPPING SYSTEM DESCRIPTION

This multi-cell T-mapping system has been developed for the Cornell ERL project which requires the $Q_0$-value of 7-cell cavities to achieve $2\times10^{10}$ at accelerating gradient 16MV/m [7]. For a 1.3GHz SRF cavity, the heating from RF causes on average about a 10mK temperature rise on the exterior wall at 25MV/m in a 2K helium bath [9]. Thus the resolution of this T-map is required to be approximately 1mK.

### Thermometers and boards

The Cornell multi-cell T-map has nearly two thousand thermometers. The temperature sensor is a 100Ω carbon Allen-Bradley resistor (5% 1/8 W). Carbon is a semiconductor, its resistance increases exponentially when temperature drops. Figure 1 shows the schematic of sensor and its picture. Dow Corning vacuum grease was applied on the varnished side of the thermometers prior to inserting the boards in cages. Silicone-based Dow Corning grease has similar thermal conductivity to traditional APIEZON grease in superfluid helium, but much cheaper than APIEZON grease. When the thermometers press to the cavity wall, the grease spreads into remaining gaps and prevents superfluid helium from cooling the sensors. The sensor is able to detect the temperature rise of the cavity wall with 25% efficiency [3].

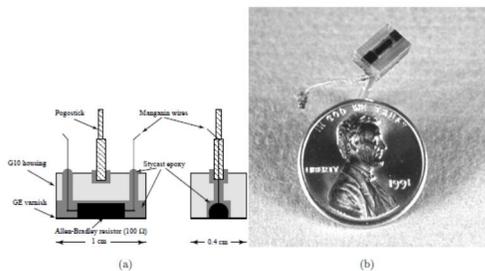

Figure 1: (a) Thermometer schematic, (b) Image of a thermometer. [3]

Two sets of 3-cell boards and one set of 1-cell boards were used for 7-cell SRF cavity. A 3-cell board has two channels addressed to the sensors for 3-cells. The sensor boards are spaced azimuthally every 15° around the cavity; totally it has 24 boards around the azimuth. Each cell is covered by an 11×24 thermometer array. Figure 2 shows a 3-cell and a 1-cell board. The quantity of thermometers for the whole 7-cell cavity is 1848.

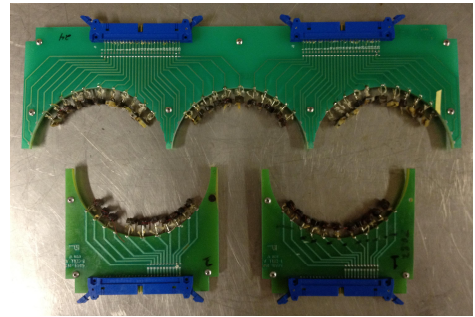

Figure 2: A 3-cell board and 1-cell boards

### Insert for cryogenic test pits

Cornell constructed a new test insert for hanging the multi-cell cavity with full T-map. In figure 3 the left image shows the Cornell multi-cell T-map attached on middle 7-cells of a 1.3GHz 9-cell cavity; the right image is a picture showing several boards removed to expose the cavity. The insert has a feedthrough on the top-plate for extracting T-mapping cables out of the Dewar. A coupler motor and Ion pump were built on the insert as well, which allows us to adjust the coupler position as well as to keep the cavity under high-vacuum during the RF test. Three well-calibrated Cernox thermometers were attached on the insert for T-map calibration.

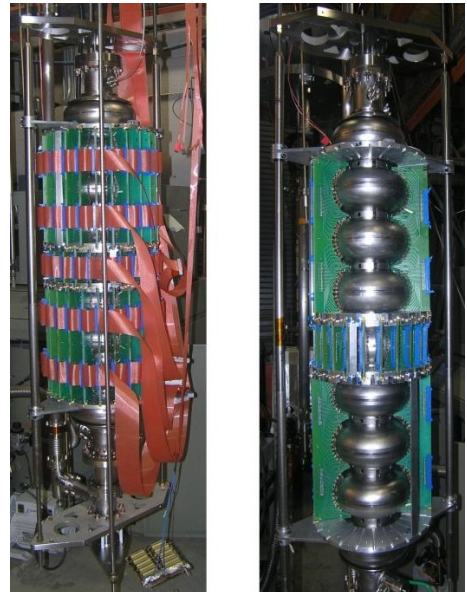

Figure 3: Picture of cryogenic insert

### Electronic and Scan Scheme

The traditional scan-scheme for single-cell T-map has each sensor connected to two wires. A multiplexer scans each sensor one by one. Therefore the number of wires is two times the number of sensors. However, for a multi-cell T-map application the wire numbers would reach approximately four thousand if that single-cell T-mapping scheme were adopted. A simple scheme was proposed [9] to reduce wire numbers. Figure 4 illustrates the schematic of a new scan scheme for a multi-cell T-map.

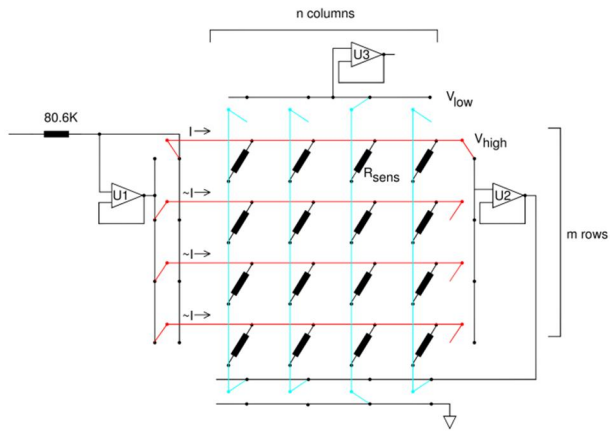

Figure 4: Schematic of new scheme for multi-cell T-map

In the conceptual view of Figure 4, the working principle of the multi-cell scheme is shown. Each of an array of $m \times n$ resistive thermal sensors has one lead attached to a "column" wire, and one lead attached to a "row" wire. In the case of our implementation, there are 24 columns and 11 rows in each array, and our measurement system can process the results from up to 6 such arrays simultaneously. This gives a potential capability for 3600 sensors with about 300 wires going into the cryostat. Each column wire may be connected by a CMOS switch to an operational amplifier (U3), and by a low-on-resistance CMOS DPDT switch to either ground or to the output of op-amp U2. Likewise, each row wire may be connected by a CMOS switch to the input of the follower op-amp U2, and through a low-on-resistance CMOS DPDT switch to either a current source or to the output of op-amp U2 which tracks the output voltage of the current source. In a measurement of a particular sensor, the column wire for that sensor is connected to U3 and the system ground, while the other column wires are connected to the output of U2, while the row wire for that sensor is connected to the current source and the input of U2 and the output of U1. Thus all the rows and all but one of the columns are at the same potential (neglecting the wiring resistance), and a current given by the current flows through the selected resistor. There is a current of similar magnitude, though depending in detail on the resistance of the other sensors, through all of the other sensors on this same column. The resistance of the selected sensor is determined by the voltage difference between the output of U2 and U3. It is necessary to keep the measurement voltage relatively small to avoid overheating of the sensor by the measurement power. We use a current source providing about 4 microamperes through the nominal 10K sensor resistance at the operating temperature, giving approximately 40mV of "typical" signal across a sensor. To get meaningful readings, the op-amps used in the circuit must be very low in voltage offset (ca. 1 microvolt), or else the unintended currents through the other resistors in the array will cause unacceptable errors. In our case, we use 5×IDC-50-wire cryogenic ribbon cables for total 1848 thermometers.

The analog sensing and digital control for the signal processing board is provided by a National Instruments PXI-1033 chassis. In the chassis is a PXI-6123, 8-channel analog input, 500 kHz DAQ module, and a PXI-6509, 96 DIO module for thermometers addressing, signal conditioner gain select, carrier waveform select, and reference thermometer control. A Matlab code on a Windows PC with a PCI slot controls the PXI chassis.

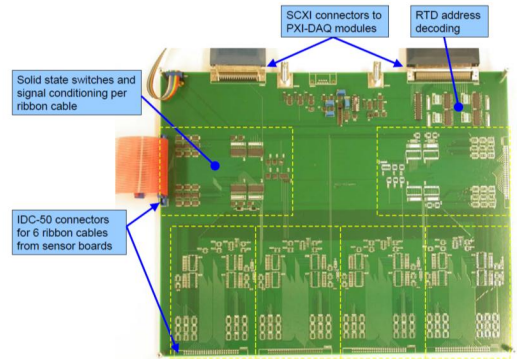

Figure 5: Picture of signal processing board

The switching and signal processing board fabricated for multi-cell cavity is shown in Figure 5. It is an 8-layer board with surface-mount components on both sides. There is a section for decoding the thermometer address and 6 duplicated channels to process the signals from 6 ribbon cables.

## CRYOGENIC TEST

We tested a 1.3GHz 9-cell cavity which is TESLA shape with the T-mapping system. The cavity was vertically electro-polished about 5μm, followed by high pressure water rinsing and clean assembly in a class 10 clean room. No 120°C baking was applied on the cavity.

### Calibration and Sensitivity

A well-calibrated Lakeshore Cernox thermometer is utilized as a temperature reference for the helium bath during cooling down from 4.2K to 1.6K. Figure 6 is a sensor's calibration curve. The X-axis is the resistance value of the carbon resistor, and the Y-axis is the reciprocal of the temperature measured by the Cernox sensor.

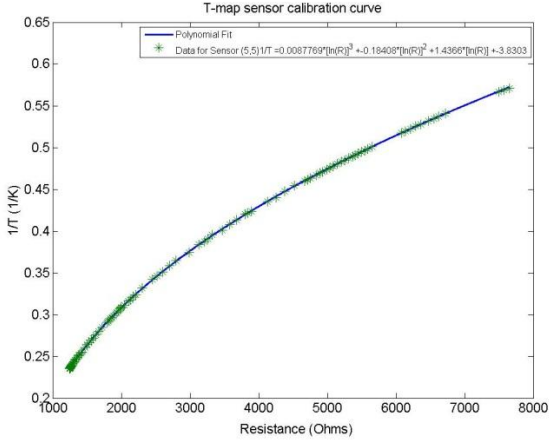

Figure 6: 1/T vs. Resistance curve during cooling down.

A polynomial function was adopted to fit the curve, shown in equation (4).

$$\frac{1}{T} = a_n x^3 + b_n x^2 + c_n x + d_n$$
$$x = ln(R)$$
(4)

Here $T$ is the bath temperature, $R$ is the resistance of the carbon resistor, $a_n$, $b_n$, $c_n$, and $d_n$ are fitting parameters for sensor $n$. To avoid measurement error caused by bath temperature variation, the data-scan program measures bath temperature right after each T-mapping sensor measurement. Therefore each Allen-Bradley resistor has an individual calibration curve and fitting parameters. Each resistor's thermal performance is slightly different from one to another.

By using fitting curves, it is possible to calculate $\frac{dR}{dT}$ which represents the sensitivity of each T-mapping sensor. An Allen-Bradley resistor value is about $12k\Omega$ at 1.6K, $\frac{dR}{dT}$ is approximately 30 Ω /mK; and $\frac{dR}{dT}$ is about 10 Ω /mK at 2K.

*Noise level analysis*

Noise comes from the environment and the electronic system. Good grounding of all instruments helps to reduce noise from environment. We use a grounding line to connect the signal generator, RF amplifier, T-mapping electronics, and Dewar together with a building grounding point.

Increasing measurement sampling number is an effective way to reduce noise from the electronic system. The PIXI controller can sample the sensor voltages at a 500 KHz rate, and average readings for $2^N$ samples. The $N$ is the parameter for setting sampling number in the Matlab program. The program returns average value of each thermometer. Figure 7 shows the standard deviation of resistance decreasing with increasing sampling number. Each point in figure 7 was calculated from 10 scans at different $N$ value. The dash line in the plot is a theoretical curve which is the standard deviation versus $\sqrt{1/2^N}$.

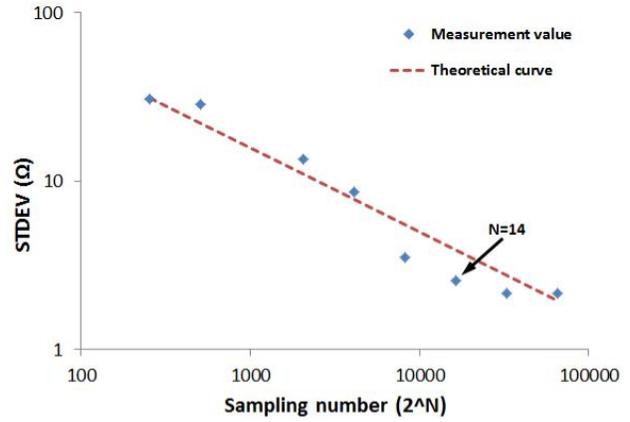

Figure 7: Standard deviation vs. sampling number

The standard deviation is reduced to about 2Ω, when $N$ is set to 14, corresponding to a sampling number of about 16000. We don't choose a higher $N$ value because higher $N$ will significantly extend scan time. The noise from the electronic system is about 200μK in a 2K helium bath; and 67μK in 1.6K helium. The total scan time is about 100s.

Thermal conductivity of grease as well as press-force of sensor against cavity-wall affects T-map noise level as well. Figure 8 is a lateral view of the subtraction between two T-map scans without applying RF field in 2K Helium bath. The total noise of the T-map system is about 1mK which achieved our initial goal. Work continues on improving temperature stability of the helium bath, which may be the present limit on system temperature noise.

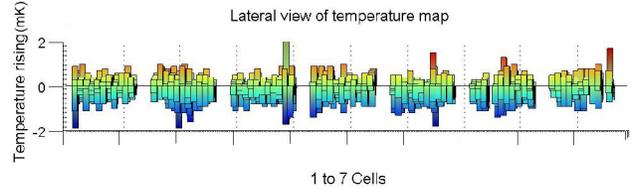

Figure 8: 3-sigma noise level of T-map system. The one sigma noise is only about 0.2mK.

*Testing results of T-map*

The 9-cell cavity A9 was tested in a 2K helium bath and quenched at 21MV/m in pi-mode. The T-mapping boards covered the middle 7 cells of the cavity. Temperature-map data was taken at an accelerating gradient very close to quench. Figure 9 depicts the temperature-rise ($\Delta T$) map. The T-mapping result is a 3D bar plot in which the x-axis and y-axis are the coordinates of T-mapping sensors and the z-axis is $\Delta T$. In Figure 9, the left plot is the top view of the T-map result, each blue rectangular area represents a thermometer array attached on a cell, and has same sequence with cavity cells from top to bottom; the $\Delta T$ is showed by colours. The right plot is an isometric view of the T-map with a finer scale (0-15mK) of $\Delta T$ which better displays tiny heating on cells.

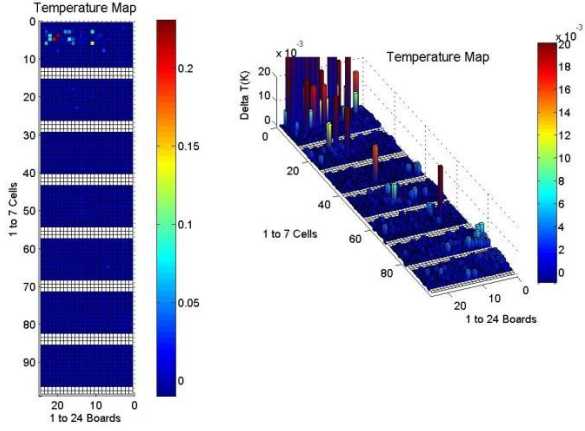

Figure 9: Map of temperature increase for 7-cells of an ILC cavity. The dominantly heated cell at fields close to quench is clearly visible.

Most heating-spots were found on the top cell which is the top second cell of the 9-cell cavity. The highest $\Delta T$ is about 0.2K. Heating is expected to be a function of the magnetic field squared, as shown in equation (5)

$$P_{loss} = \int_A \frac{1}{2} R_s |\vec{H}|^2 dA = \gamma \sum_{n=1}^{1848} C(\Delta T_n A_n) \quad (5)$$

$$\gamma = \frac{\int_{A\,tot} |\vec{H}|^2 dA}{\int_{A\,measure} |\vec{H}|^2 dA} \quad (6)$$

$$C = \frac{P_{loss}}{\gamma \sum_{n=1}^{1848} \Delta T_n A_n} \quad (W/(Km^2)) \quad (7)$$

$$P_{loss} = \frac{(E_{acc} L)^2}{R/Q \; Q_0} \quad (8)$$

Here $P_{loss}$ is the total power loss on the surface, $A$ is the area of cavity surface, $R_s$ is the surface resistance, $\vec{H}$ is the surface magnetic field, $\Delta T$ is the temperature rise. As the T-map sensors cover the high-magnetic field region of the cells, the coefficient $g$ is the ratio of the total power and the power measured by the T-map. The coefficient $C$ is determined by equation (7), where $A_n$ is the area of the region closest to thermometer $n$, $DT_n$ is the temperature rise detected by thermometer $n$. The $P_{loss}$ in equation (7) is calculated by equation (8), where $E_{acc}$ is the accelerating gradient, $L$ is the length of a 9-cell cavity, $\frac{R}{Q}$ is the ratio of shunt impedance and quality factor which is a constant determined by cavity shape, $Q_0$ is the quality factor, The coefficient $C$ indicates temperature response of the thermometers to a unit of power flux. It is assumed that $C$ is approximately the same for all T-mapping sensors. $E_{acc}$ and $Q_0$ are measured from RF [10]. Hence with the equation (7) and (8), the $C$ is possible to be calibrated at the highest accelerating gradient and be applied to all the other field level.

Figure 10 is the summary of $\Delta T$ versus $H^2$ curves; the axis on the right is the $Q_0$ of the cavity. Equation (9) depicts the $\Delta T$ and $H^2$ relationship of the sensor $n$. Here $R_{s\,n}$ is the surface resistance close to the sensor $n$, $\vec{H}_n$ is the surface magnetic field at the sensor $n$, as the T-map sensors only cover the high magnetic-field region of each cell, thus $\vec{H}_n \approx H_{peak}$.

$$\Delta T_n = \frac{1}{2C} R_{s\,n} \vec{H}_n^2 \quad (9)$$

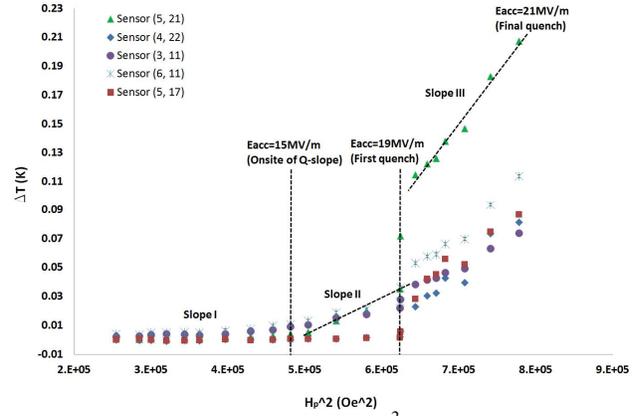

Figure 10: $\Delta T$ vs. $H^2$ Curve

The slopes of $\Delta T$ versus $H^2$ curves indicate $R_s$ at the sensors from equation (9). Three slopes were observed. The slope I started from an acceleration gradient from 0 to 15MV/m. The surface resistance was consisted by BCS resistance and residual resistance. The slope II started at an accelerating gradient ($E_{acc}$) 15MV/m where the onset of Q-slope is. The surface resistance was increased due to the Q-slope. The heating increased dramatically around 19MV/m at the Multipacting barrier of the TESLA-shape cavity [11]. The Multipacting triggered the first quench and processing of the cavity. After the Multipacting had been processed out, the heating increased with $H^2$ by the slope III until hard quench (21MV/m). The slope III indicates the surface resistance was increased further by flux trapping after the first quench.

The hot-spot locations are close to the electron-beam-welding seam, depicted in Figure 11 which is a detailed view of cell 1. The three hottest regions, for which $\Delta T$ is larger than 0.1K, are marked in Figure 11. And there are many small heating spots ($\Delta T$ <0.1K) surrounding them.

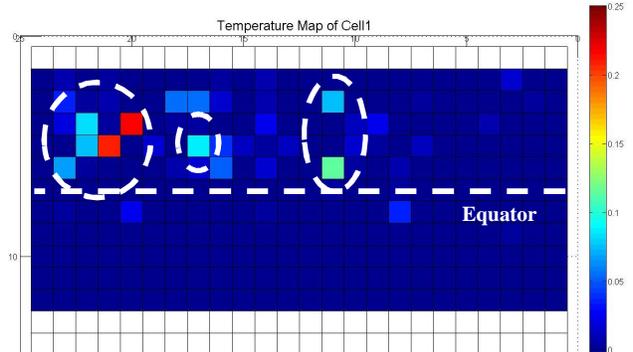

Figure 11: Detail T-map view of the dominantly heated Cell 1

*Converting $\Delta T$ to power loss and quality factor*

From equation (5) we know that the power loss on the cavity surface can be calculated from $DT$. Equation (10) gives a formula of power loss by summing all $DT$.

$$P_{loss}^{Cal} = \gamma \sum_{n=1}^{1848} C(\Delta T_n A_n) \qquad (10)$$

Here the coefficients $C$, $A_n$, and $\Delta T_n$ are the same parameters in equation (5) and (6). The $Q_0^{Cal}$ calculated from $P_{loss}^{Cal}$ is given by

$$Q_0^{Cal} = \frac{\omega U}{P_{loss}^{Cal}} = \frac{\frac{(E_{acc}L)^2}{R/Q}}{P_{loss}^{Cal}} \qquad (11)$$

Here $w$, $U$, $E_{acc}$, $L$, $\frac{R}{Q}$ are the same parameters as in equations (8). The calculation result is shown in Figure 12. The left y-axis is the quality factor $Q_0$ versus accelerating gradient $E_{acc}$ curve and the right y-axis is $P_{loss}$ vs. $E_{acc}$ curve. The calculated $Q_0$ and $P_{loss}$ are compared with the measurement result from the RF.

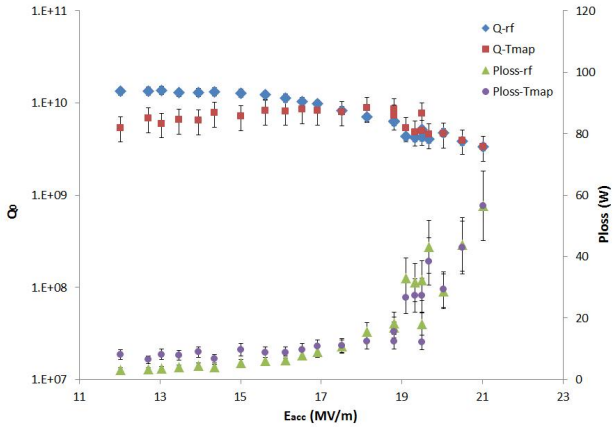

Figure 12: $Q_0$, $P_{loss}$ comparison between calculated from T-map and measured from RF.

The measurement curves match the calculated curve when the accelerating gradient is higher than 15MV/m where the heating started. Below 15MV/m, the heating from the cavity is too small to be detected accurately.

By summing the temperature rising of individual cells, we obtain power loss and quality factor of each cell. It is given by equations (12) and (13) which are modified from equations (10) and (11).

$$P_{loss}^{Cell} = \gamma \sum_{n=1}^{264} C(\Delta T_n A_n) \qquad (12)$$

$$Q_0^{Cell} = \frac{\omega U}{P_{loss}^{Cal}} = \frac{\frac{1}{9}\frac{(E_{acc}L)^2}{R/Q}}{P_{loss}^{Cal}} \qquad (13)$$

Figure 13 shows the power loss of individual cells and the total $P_{loss}$ of the cavity. Figure 14 illustrates the quality factor of individual cells and the total $Q_0$ of the cavity. In pi-mode, each cell has even stored energy and field distribution. Thus it is supposed to have even power loss on surface of cells in the ideal case. However, the results from T-map indicate the measured cell 1 generated the most power loss on the surface, and degraded the total $Q_0$ starting from 15MV/m. The equation (3) suggests that the total $Q_0$ of the cavity is limited by the Q-value of the worst cell. The cavity test results also indicate that the Q-slope is possible to be a localized effect, and the onset of the Q-slope happened lower than the normal case which is supposed to be 20-25MV/m.

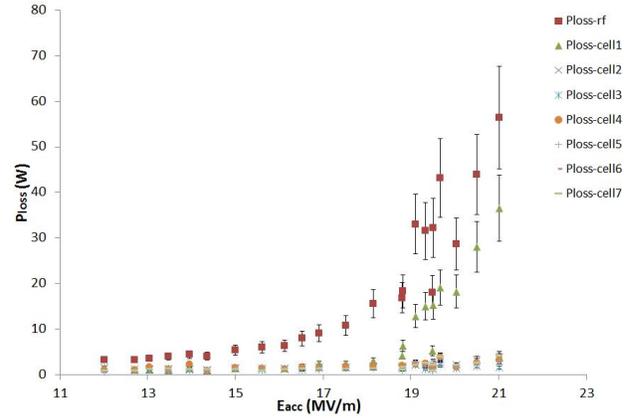

Figure 13: The power losses in individual cells, adding up to the loss of the full cavity

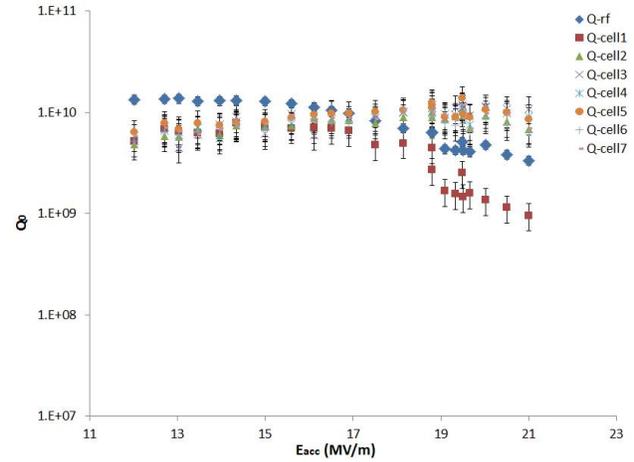

Figure 14: The quality factor of individual cells compared to total $Q_0$

Figure 15 depicts how the hot-spots affect the Q-value of the cavity. In the figure, the red square marker (■) is calculated from the average surface resistance of cold region. In other words, the calculation excludes the hot-spots which temperature increases are higher than 50mK. The blue diamond marker (◆) is calculated from the average surface resistance including the hot-spots. The difference of the Q-value at the accelerating gradient 21MV/m is $2.2\times10^9$. The Q-value would increase 1.65 times higher if the cavity were hot-spots ($\Delta T >50mK$) free. The curves (▲) and (●) are the comparison of power loss with and without the hot-spots. At 21MV/m, 22.5W power lost on the hot-spots region, which is about 40% of the total power losses.

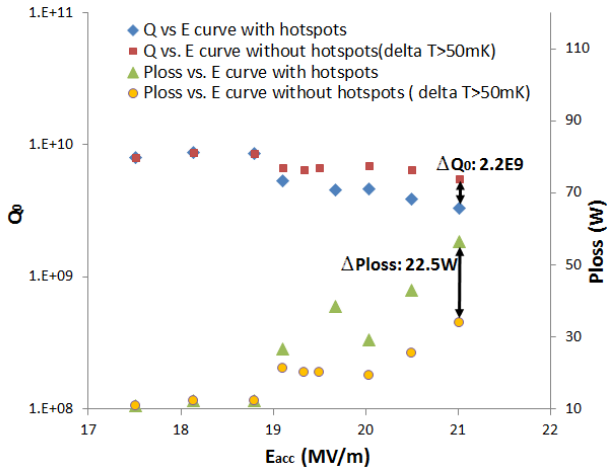

Figure 15: The comparison of the $Q_0$ vs. $E_{acc}$ curves with and without the hotspots

*Surface resistance map*

A surface resistance map is calculated from the *DT* of each thermometer, which is given by equation (14) deduced from equation (9).

$$R_{s\,n} = \frac{2C\Delta T_n}{H_{peak}^2} \quad (14)$$

Figure 16 shows a surface resistance map when the cavity was at 21MV/m, close to quench.

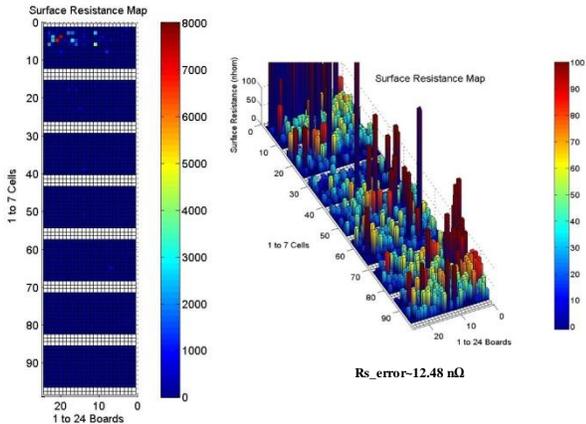

Figure 16: Map of the surface resistance $R_s$

The maximum $R_s$ value is close to 8000nΩ, and the average value of all $R_s$ map is 74.26nΩ which agrees with the value (80.84nΩ) calculated from equation (1) by using the geometry factor and $Q_0$. The error of the surface resistance is about 16.8% calculated from the measurement error of $\Delta T$, accelerating gradient ($E_{acc}$) and Q-value by uncertainty propagation formulas.

It should be pointed out that the surface resistance calculated from T-map combines the effects of all types of surface heating such as BCS surface resistance heating, residual resistance heating, field emission, flux trapping, and so on.

## THE USAGE OF THE T-MAP SYSTEM

Except to detect temperature increases as well as to obtain surface resistance map, the T-map system is able to be used for the quench location detection. Unlike the OST system [12], the T-map system enables to detect pre-quench signals and distinguish a global or local quench. The information would help for quench mechanism researches.

The T-map data is very useful information which gives an effective feedback of cavity surface preparations. Therefore with large statistics of T-map data, it's possible to establish a relationship between each surface-preparation step and the cavity performance. It would improve the cavity preparation procedures, thus definitely will push the cavity performance yield up.

The hot-spots information can be used for avoiding the cavity surface damage from a heavy field mission as well as for adding localized cooling on the hot-spots.

## CONCLUSION

A new multi-cell T-map system has been constructed at Cornell University. The system has nearly two thousand sensors to cover 7-cell SRF cavities for the Cornell ERL project. A new scan-scheme was adopted to reduce wire numbers. The resolution of temperature achieved is 1mK. A 9-cell SRF cavity (A9) was tested with the T-map system. By converting $\Delta T$ to power loss and quality factor, we found the first cell generated most heating and degraded cavity $Q_0$. A surface resistance map was obtained from the temperature map. Hot-spots were identified and it was found that the hot-spots consumed 40% of total energy and degraded $Q_0$ about 1.65 times.

## ACKNOWLEDGMENT

I am grateful to James Sears, John Kaminski and Matt Rifenburg for their mechanical support. I would like to thank John Barley and Lenard Hirshman for his support of electronics.  I am also grateful to Vivian Ho for her support of the programming. E. Chojnacki provided the initial electronics design, he produced the electronic boards, and the initial version of our control program. Without his initiative this project would not have been possible.